%% file: root.tex
\documentclass[letterpaper, 10 pt, conference]{ieeeconf}  

\IEEEoverridecommandlockouts                              

\usepackage{amsmath,amssymb,amsfonts}
\usepackage{algorithmic}
\usepackage{graphicx}
\usepackage{algorithm,algorithmic}
\usepackage{textcomp}
\usepackage{float}
\usepackage{booktabs}
\usepackage{svg}
\usepackage{pdfpages}
\usepackage{ragged2e}
\usepackage{comment}
\usepackage{makecell}   
\usepackage{array}      
\usepackage{multirow}
\usepackage{booktabs}
\usepackage{tabularx}
\usepackage{threeparttable}

\usepackage{cite}
\makeatletter
\let\NAT@parse\undefined
\makeatother
\usepackage{hyperref}

\newcolumntype{Y}{>{\raggedright\arraybackslash}X}

\usepackage{acronym}
\input{acronyms}

\newcommand{\review}[1]{\textcolor{black}{#1}}

\title{A Wearable Multimodal Ultrasound+Inertial System for Real-Time Virtual Reality Interaction}

\author{Giusy Spacone$^{1*}$, Sebastian Frey$^{1*}$, Enzo Baraldi$^{1}$, Mattia Orlandi$^{2}$, Luca Benini$^{1,2}$, and Andrea Cossettini$^{1}$
\thanks{$^{1}$G. Spacone, S. Frey, E. Baraldi, L. Benini, and A. Cossettini are with the Integrated Systems Laboratory of ETH Z{\"u}rich, Z{\"u}rich, Switzerland (\texttt{gspacone@iis.ee.ethz.ch}).}
\thanks{$^{2}$M. Orlandi and L. Benini are with the Department of Electrical, Electronic and Information Engineering, University of Bologna, Bologna, Italy.}
\thanks{*G. Spacone and S. Frey are first co-authors.}
}

\makeatletter
\def\ps@IEEEtitlepagestyle{%
  \def\@oddhead{}%
  \def\@evenhead{}%
  \def\@oddfoot{%
    \hfil
    \parbox{0.95\textwidth}{%
      \centering\scriptsize
      © 2026 IEEE. Personal use of this material is permitted. Permission from IEEE must be obtained for all other uses, in any current or future media, including reprinting/republishing this material for advertising or promotional purposes, creating new collective works, for resale or redistribution to servers or lists, or reuse of any copyrighted component of this work in other works.
    }%
    \hfil
  }%
  \def\@evenfoot{}%
}
\makeatother

\begin{document}

\maketitle
\thispagestyle{empty}
\thispagestyle{IEEEtitlepagestyle}

\input{sections/0_Abstract}

\input{sections/1_Intro}
\input{sections/2_Related}

\input{sections/3_Methods}
\input{sections/4_Results}

\input{sections/5_Conclusions}
\vspace{-0.1cm}
\vspace{-0.1cm}

\bibliographystyle{IEEEtran}
\bibliography{biblio}

\end{document}

%% file: acronyms.tex
\acrodef{EMG}[EMG]{electromyography}
\acrodef{sEMG}[sEMG]{surface EMG}
\acrodef{US}[US]{ultrasound}
\acrodef{AUS}[AUS]{A-mode ultrasound}

\acrodef{ML}[ML]{machine learning}
\acrodef{DL}[DL]{deep learning}
\acrodef{RMS}[RMS]{root mean square}
\acrodef{MAV}[MAV]{mean absolute value}
\acrodef{WL}[WL]{waveform length}
\acrodef{ANN}[ANN]{artificial neural network}
\acrodef{CNN}[CNN]{convolutional neural network}
\acrodef{BPTT}[BPTT]{backpropagation-through-time}
\acrodef{TSNE}[TSNE]{T-distributed stochastic neighbor embedding}

\acrodef{TAC}[TAC]{Target Achievement Control}
\acrodef{SR}[SR]{Success Rate}
\acrodef{CR}[CR]{Completion Rate}
\acrodef{CT}[CT]{Completion Time}
\acrodef{SE}[SE]{Stability Error}

\acrodef{HMI}[HMI]{Human-Machine Interface}

\acrodef{IMU}[IMU]{inertial measurement unit}
\acrodef{ACC}[ACC]{accelerometry}
\acrodef{SoA}[SoA]{state-of-the-art}
\acrodef{VR}[VR]{virtual reality}

\acrodef{HGR}[HGR]{Hand Gesture Recognition}
\acrodef{DoF}[DoF]{degree-of-freedom}
\acrodefplural{DoFs}[DoFs]{degrees-of-freedom}
\acrodef{OC}[OC]{Open-Close}
\acrodef{Pro-Sup}[Pro-Sup]{Prono-Supination}
\acrodef{Fle-Ext}[Fle-Ext]{Flexion-Extension}
\acrodef{RUD}[RUD]{Radial-Ulnar Deviation}
\acrodef{CH}[CH]{Channel}
\acrodefplural{CHs}[CHs]{Channels}

\acrodef{BLE}[BLE]{Bluetooth Low Energy}

%% file: sections/0_Abstract.tex
\begin{abstract}
A-mode ultrasound (US) is a promising sensing modality for Virtual Reality (VR) interaction, as it enables the mapping of muscular activity into control commands while retaining the benefits of wearable sensing. However, existing approaches still face limitations in terms of wearability and interaction complexity, often relying on external hardware such as cameras. In this work, we propose a fully wearable multimodal interface for real-time VR-interaction, based on concurrent US and inertial (accelerometry) sensing from the forearm and upper arm. The system is built on the WULPUS platform and integrates an end-to-end software framework for real-time acquisition, visualization, and communication with a Unity-based VR environment. A multimodal learning pipeline is introduced for concurrent hand pose and forearm position estimation in 2D space. 
The interface is evaluated through offline and online experiments with five subjects, during the execution of three functional tasks: cylinder grasping (gross motor) and relocation, marble pinching (fine motor) and relocation, and liquid pouring. 
For offline experiments, we collect 5 acquisition sessions across multiple days, achieving an average inter-session accuracy across subjects of 80$\pm$6\% for hand pose estimation and 77$\pm$7\% for forearm position estimation. Online validation with minimal fine-tuning (5 min) demonstrates success rates of 92.0$\pm$16.0\%, 88.0$\pm$9.8\%, and 96.0$\pm$8.0\% for the three tasks, respectively. With a power consumption of only 19.9~mW, our system enables more than 2.5 days of continuous use on a small 350 mAh LiPo battery without the need for recharge, enabling truly wearable, multimodal, and functionally meaningful VR interaction.  
\end{abstract}


%% file: sections/1_Intro.tex
\vspace{-0.2cm}
\section{Introduction}\label{sec:intro}
Hand gesture recognition and arm position control are key enablers of modern \acp{HMI}, particularly in \ac{VR} environments where interaction with virtual content requires natural and continuous hand input \cite{nguyen_2023_handvrreview}. \ac{VR} systems are increasingly used in applications ranging from rehabilitation and prosthetic training \cite{peral_2022_emgvrreview} to consumer devices\review{,} such as smart glasses \cite{kaifosh2025generic}.

Wearable sensing is a common solution for continuous hand tracking without relying on cameras, which can suffer from occlusions and out-of-view conditions. 
Existing wearable approaches include instrumented gloves and biosignal-based systems. Glove-based systems capture detailed finger kinematics but are often cumbersome and require frequent calibration  \cite{rodriguez_2021_gloves_review}. Biosignal-based systems instead estimate hand activity from the forearm and/or upper arm neuromuscular activity. In this context, \ac{EMG} has been the predominant sensing modality for \ac{HGR} \cite{shin_methodologic_2024}. More recently, \ac{US} sensing has emerged as a promising alternative, as it provides deeper signal penetration and higher spatial resolution \cite{yang_2024_usneuroroboticreview}. Unlike conventional B-mode imaging, \ac{US} is also well suited for low-power wearable acquisition. In addition, recent research reported the benefit of sensor fusion involving \ac{US}, with a main focus on its combination with \ac{EMG} \cite{yin_2025_multisensor, spacone_2025_emgus}.

Despite these demonstrated advantages of \ac{US}-sensing, most evaluations of gesture recognition are performed offline. This represents a limitation, as offline analysis does not necessarily translate into robust online operation \cite{yang_2024_usneuroroboticreview}. Existing online evaluations either address simple tasks (such as the \ac{TAC} test), or target more complex goals (such as \ac{VR} interaction) with hardware that is not fully wearable \cite{sgambato_2025_virtual}, which limits user mobility and practical usability. Furthermore, previous works primarily focused on controlling hand or wrist states, while realistic manipulation tasks also require estimation of hand and forearm position.


In this work, we address these challenges by proposing a fully wearable multimodal sensing solution based on US and accelerometry for simultaneous estimation of hand pose and forearm position, integrating the wearable sensing platform with a real-time VR online interaction environment. The main contributions of the work are:
\vspace{-0.1cm}
\begin{itemize}
    \item A multimodal wearable sensing system combining \ac{US} and \ac{ACC} data for concurrent monitoring of the forearm and upper arm muscular states, based on the WULPUS platform \cite{frey_wulpus_2022}\footnote{\url{https://github.com/pulp-bio/wulpus}}, coupled with an end-to-end, open-source real-time framework for data acquisition and \ac{VR} interaction\footnote{\url{https://github.com/pulp-bio/biogui}}.
    
    \item A multimodal learning pipeline for concurrent hand pose (6 classes) and forearm position (3 classes) estimation across three functional tasks, with an average offline inter-session accuracy across five subjects of $80\pm6\%$ and $77\pm7\%$ for hand pose and forearm position, respectively.
    

    \item A real-time validation across three functional tasks (Cylinder Grasping and Relocation, Marble Pinching and Relocation, and Liquid Pouring), including an analysis of the fine-tuning data required to address the challenges caused by sensor repositioning. With only 3 fine-tuning repetitions, we obtain task \ac{SR} values of \(92.0\pm16.0\%\), \(88.0\pm9.8\%\), and \(96.0\pm8.0\%\), and corresponding task \ac{CT} values of \(10.47\pm4.73\)\,s, \(10.83\pm2.02\)\,s, and \(8.00\pm1.56\)\,s for Cylinder Grasping and Relocation, Marble Pinching and Relocation, and Liquid Pouring, respectively. 

\end{itemize}

%% file: sections/2_Related.tex
\vspace{-0.2cm}
\section{Related Works}\label{sec:related}
\vspace{-0.1cm}
This section summarizes state-of-the-art works in real-time wearable \acp{HMI} based on \ac{US}, with a focus on online control and functional interaction. A summary is provided in Table \ref{tab:us_comparison_transposed} (discussed in more detail in Sect.~\ref{soa:comparison}).

Early studies demonstrated online recognition of discrete hand and finger states \cite{yang_2019_towards,lu_2022}. While these works showed the feasibility of real-time \ac{US}-based classification, they focused on discrete gestures and did not evaluate functional interaction tasks.
%
Subsequent work overcame this limitation by proposing virtual \ac{TAC} tests \cite{yang_2022_sonomyographic}, where users move a virtual hand into a target posture via online simultaneous and proportional control of two \acp{DoF} at the hand and wrist, with performance evaluation based on \ac{CR}, motion \ac{CT}, and \ac{SE}.
%
Extensions of this paradigm included the simultaneous control of three \acp{DoF} to interact with a cursor within a 2D grid \cite{sgambato_2023}.
Despite moving beyond static offline analysis, these works share several limitations: online evaluation is performed immediately after data collection, without considering transducer repositioning; the \ac{TAC} protocol remains a simplified target-achievement task; in addition, the hand configuration is evaluated under fixed spatial conditions, limiting the execution of sequential manipulation tasks with spatial awareness. As discussed in \cite{yang_2024_usneuroroboticreview}, online protocols should instead prioritize daily life activities involving movement and posture changes.
Recent work by Sgambato et al. \cite{sgambato_2025_virtual} addressed part of these limitations by implementing model training using offline-collected data from ten subjects, without subject-specific fine-tuning prior to online validation. The proposed virtual environment includes a prosthetic hand, used to perform five interaction tasks with virtual objects, and \ac{US} data are used to control four \acp{DoF} at the hand and the wrist. Unlike previous \ac{TAC}-based studies, the authors do not report \ac{SR} or \ac{CT} metrics; instead, they compare the predicted joint angles against those measured by an optical tracking system used as ground truth.
However, \ac{US} is still used to control only hand poses: the hand position and orientation within the environment are obtained from an external optical motion capture system.
Using distinct sensing units on the forearm and upper arm was suggested as possible improvement \cite{sgambato_2025_virtual}; this configuration was previously investigated by Tang et al. \cite{tang_2025_synchronous} with $3\times3$ transducer arrays placed on the forearm, upper arm, and chest wall to perform concurrent hand gesture estimation and forearm-upper arm kinematic regression for the control of a 4-\ac{DoF} robotic hand, however, using a non-wearable setup.

Overall, these studies highlight the potential of \ac{US}-based interfaces for \ac{HMI}. Remaining open challenges include evaluating robustness to sensor repositioning, support for spatially meaningful multi-step manipulation tasks, and the development of fully wearable solutions capable of jointly estimating hand state and upper limb configuration. In this work, we address these limitations by proposing \textit{a fully wearable, distributed, multimodal \ac{US} and \ac{ACC} data interface for simultaneous hand pose estimation and spatial forearm position control, exploiting two sensing units placed on the forearm and upper arm to enable real-time VR object interaction}.

%% file: sections/3_Methods.tex
\vspace{-0.4cm}
\section{Methods}\label{sec:Methods}
\vspace{-0.1cm}
This section describes the proposed system and the experimental methodology. We first present the wearable acquisition setup for concurrent \ac{US}--\ac{ACC} sensing from the forearm and upper arm. We then describe the software architecture enabling real-time data acquisition, visualization, and communication with a \ac{VR} environment. Next, the system is used to collect an offline task-oriented dataset, using a \ac{CNN}-based model for hand pose and forearm position estimation. Finally, we describe the online evaluation protocol used to assess the proposed interface during three real-time functional object-interaction tasks.

\begin{figure}[b]
    \centering
    \vspace{-0.5cm}
    \includegraphics[width=0.45\textwidth]{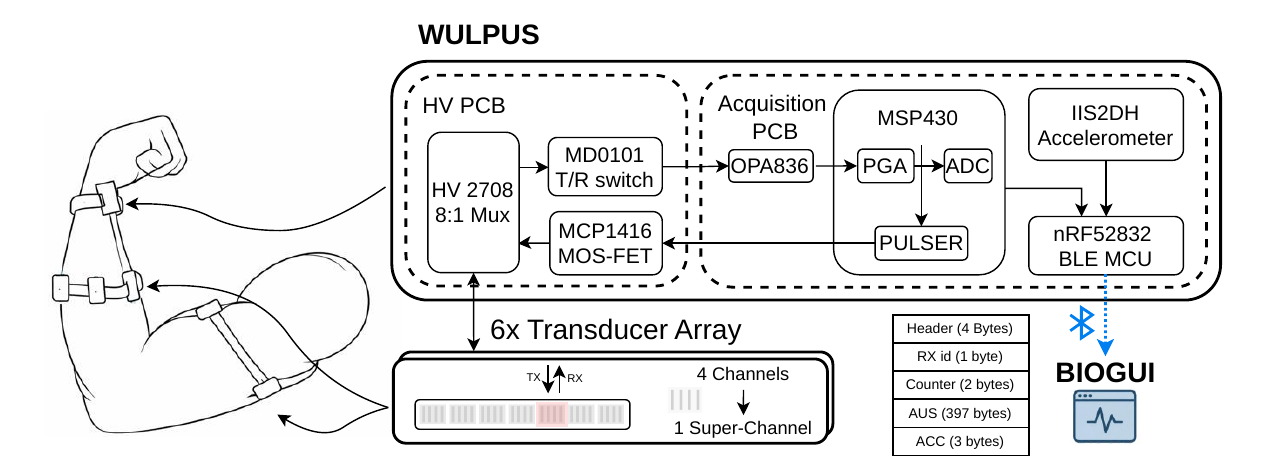}
    \vspace{-0.3cm}
    \caption{Overview of Sensing System. The acquisition electronics (with the accelerometer) is placed on the wrist. 4 US transducers are placed on the Forearm; two US Transducers are placed on the Upper Arm.}
    \label{fig:form_factor}
\end{figure}

\vspace{-0.2cm}
\subsection{Acquisition System}
\vspace{-0.2cm}
\ac{US} data are acquired using the WULPUS platform~\cite{frey_wulpus_2022} (Fig.~\ref{fig:form_factor}), operating at a pulse repetition frequency (PRF) of $30\,\text{Hz}$ and interfaced with six $2.25\,\text{MHz}$, transducers (Vermon, France). Each transducer is a 32-channel linear array where the four central elements are shorted into a single super-channel configured as a receiver-transmitter (all other elements are left unconnected). Transducers are excited sequentially in a round-robin fashion. 

Four transducers are integrated into a wearable armband \cite{spacone_tracking_24} positioned around the forearm (\acp{CH} 1-4), at $\approx 1/3$ of its length distal to the elbow, with \ac{CH} 1 placed over the extensor muscles and \acp{CH} 2-4 distributed equidistantly along the circumference toward the radial (thumb) side (as in  \cite{spacone_2025_emgus}). Two additional transducers, integrated into a stretchable band, are positioned on the upper arm at the same cross-sectional level, over the long head of the Biceps Brachii muscle (\ac{CH} 5) and the Triceps muscle group (\ac{CH} 6). Hydrogel pads~\cite{hydrogel_pads} are used for acoustic coupling. WULPUS is enclosed in a PLA casing $(31\times54\times26~mm^3,~34~g)$ worn at the wrist.

In addition, we extend the WULPUS firmware to acquire data from the embedded triaxial accelerometer (IIS2DH, STMicroelectronics) embedded in its main board.  
The accelerometer operates in high-resolution mode at $30~\text{Hz}$ with a $\pm2~g$ range.
We stream data via BLE. Every packet includes one US frame (397 samples) plus 3 accelerometer samples.

\vspace{-0.2cm}

\subsection{Software Architecture} \label{subsec:software_architecture}
The software architecture extends the BioGUI framework~\cite{biogui} to support \ac{US}.
It features three functional components (shown in Fig.~\ref{fig:biogui_interface}):

\subsubsection{Data Acquisition and Visualization}
BioGUI is a Python-based, device-agnostic GUI for biosignal acquisition~\cite{biogui}. Hardware-specific communication and packet decoding are handled by an \emph{interface layer}, while the \emph{core} implements multi-threaded signal filtering, real-time visualization, recording, and data forwarding.

Compared to biosignal time series, \ac{US} also includes depth information in addition to channels and time dimensions. To support \ac{US}, we treat each transducer as an independent 2D signal. Two dynamically switchable visualization modes were added: \textbf{A-mode}, displaying the raw echo amplitude against tissue depth for each channel\footnote{it also supports overlaying raw data, filtered data, and signal envelope computed via the Hilbert transform}, and \textbf{M-mode}, showing temporal variations for each channel as a scrolling plot where depth is mapped to the y-axis and tissue reflection intensity is encoded via a grayscale colormap.

A GUI-based configuration interface allows the user to specify WULPUS measurement parameters (e.g., pulse frequency, sampling rate, RX gain, TX/RX channel configurations) and to store/load presets.
%
Upon reception, raw bytes are decoded into \ac{US} frames and accelerometer data, and a forwarding module sends the data to the \emph{middleware} over localhost TCP.

\subsubsection{Middleware Bridge}
\begin{figure}[t]
    \centering
    \includegraphics[width=0.9\columnwidth]{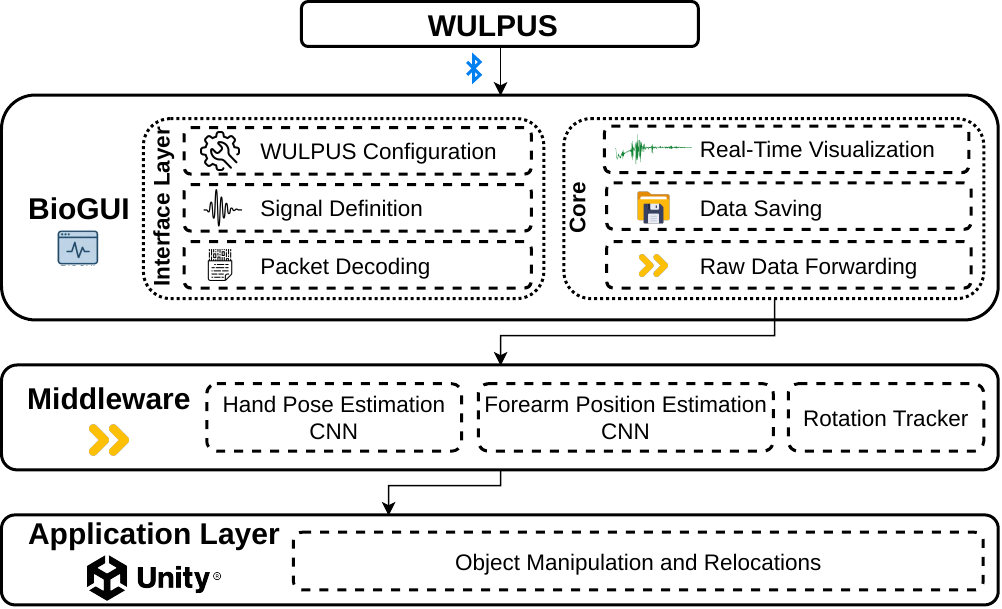}
     \vspace{-0.3cm}
    \caption{Overview of the software architecture: WULPUS communicates over BLE with the BioGUI, which forwards \acs{US} frames and accelerometer data to the \emph{middleware}; the \emph{middleware} processes the data with a CNN and sends the predictions to the application layer.}
    \label{fig:biogui_interface}
    \vspace{-0.5cm}
\end{figure}

The \emph{middleware} (Fig.~\ref{fig:biogui_interface}b) bridges the BioGUI and the application environment. It \review{handles}:\\
\textbf{\textit{i)} Hand pose classification:} using a neural network to predict hand pose from the raw \ac{US} data of all transducers together with \ac{ACC} data (more details in \review{Sec.}~\ref{subsec:architecture}). \\
\textbf{\textit{ii)} Forearm position classification:} from raw \ac{US} frames from all transducers (more details in \review{Sec.} \ref{subsec:architecture}).  \\
 \textbf{\textit{iii)} Hand Rotation Tracking:} The system performs hand rotation tracking based on the gravity vector ($\vec{g} = (x, y, z)^T$) measured by the accelerometer embedded in WULPUS and a calibrated gravity reference ($\vec{g}_{cal}$). For calibration,  the user is instructed to hold their hand extended with the palm facing downward, defining the zero-rotation point. Subsequent rotation (\ac{Pro-Sup}) is expressed relative to this reference state as $\Delta \varphi = \arctan\left(\frac{x_{cal}}{z_{cal}}\right) - \arctan\left(\frac{x}{z}\right)$, where $\Delta \varphi$ represents the angle between the calibrated gravity vector $\vec{g}_{cal}$ and the current gravity vector $\vec{g}$. \\
\textbf{\textit{iv)} Global State Estimation and Transmission:} It aggregates the hand pose and position estimations and rotation angles into a unified state packet transmitted to the {Application} environment.

\subsubsection{Application Environment}
Online validation is performed in a Unity-based \ac{VR} environment.
The environment renders a 3D scene containing an articulated virtual hand model and task-specific objects (cylinder, marble, bottle) (see Fig.~\ref{fig:task}) for real-time interaction. The virtual hand receives JSON-formatted state updates from the middleware via UDP. These updates include (i)~a discrete forearm position label (rest, forward, side), (ii)~a pose label (rest, open, close, pinch), (iii)~an \ac{ACC}-based rotation signal, and (iv)~finger-curl values set depending on the hand pose. The virtual hand is controlled as follows:

    \textbf{\textit{i)} Forearm position:} A configurable grid maps the three discrete position states received from the middleware to scene coordinates: \textit{Rest} (neutral position close to the subject's viewpoint), \textit{Forward} (in front of the hand), and \textit{Side} (laterally displaced). Transitions are smoothed with Unity SmoothDamp to avoid abrupt jumps while preserving responsiveness. \\
    \textbf{\textit{ii)} Hand Rotation:} Hand rotation (\ac{Pro-Sup}) is controlled by the \ac{ACC}-based rotation tracking algorithm described in Sec.~\ref{subsec:software_architecture}. Received rotation angles are mapped to the virtual hand axes and interpolated with per-axis angular damping. Rotation limits are enforced to match physiological constraints (supination up to $149^\circ$, pronation up to $35^\circ$).\\
   \textbf{\textit{iii)} Object interaction:} Object interaction is implemented through Unity physics and task-specific C\# logic. Each task object is equipped with a Rigid body and a collider. A spherical trigger collider surrounding the virtual hand detects objects that enter its proximity. When a grasp or pinch is detected and a compatible object is within range, a physics joint attaches the object to the hand hold point. Upon release, the joint is removed, and the object is again subject to gravity.
Different objects impose distinct grasp constraints that reflect the biomechanical requirements of the corresponding tasks:

    \textbf{\textit{i)} Cylinder:} Requires a power grasp with $90^\circ$ supination angle relative to the calibrated reference state, ensuring that the palm is vertically aligned with the object before grasping. The cylinder is used to execute gross manipulation tasks involving whole-hand object grasping and relocation, described as Cylinder Grasping and Relocation task in the rest of the manuscript. 

    \textbf{\textit{ii)} Marble:} Requires a pinch grasp (thumb–index opposition) while the forearm remains in the calibrated reference state (palm facing downward). Power grasps are rejected. This constraint reflects the precision grip typically used for manipulating small objects that require fine finger control. The marble is used to execute fine object manipulation tasks and relocation, described as Marble Pinching and Relocation task in the rest of the manuscript. 

     \textbf{\textit{iii)} Bottle:} Shares the same grasp requirement as the cylinder, i.e., a power grasp with approximately $90^\circ$ forearm supination relative to the calibrated reference state. Once grasped, the hand is automatically lifted slightly to avoid collisions with the ground plane. The bottle is used to execute pouring tasks, described as Liquid Pouring in the rest of the manuscript. It is initially filled with liquid, and the content is emptied by rotating the forearm from the supinated (vertical) position ($90^\circ$)  past the neutral position ($0^\circ$) into pronation. Pouring begins when the rotation angle crosses $0^\circ$, and the pouring rate increases linearly with the pronation angle, from a minimum at $0^\circ$ to a maximum at $45^\circ$ of pronation. The fill level drives a shader-based liquid simulation; the task is completed when the bottle is fully emptied.

For delivery-based tasks (Cylinder Grasping and Relocation and Marble Pinching and Relocation), a delivery zone is placed at the side position. Task completion is detected when the object is released within the delivery zone's trigger area. Each task follows a structured lifecycle: objects are first shown during a pre-trial countdown (visible but not interactable), then made interactable when the countdown expires, and timing begins. 
For the Liquid Pouring task, it is considered successful if it is completely emptied.
If a trial is not completed within the timeout period, it is marked as failed.
\vspace{-0.2cm}
\subsection{Offline Data Collection Protocol}\label{subsec:offline_datacollection}
\begin{figure}[t]
    \centering
    \includegraphics[width=0.8\linewidth]{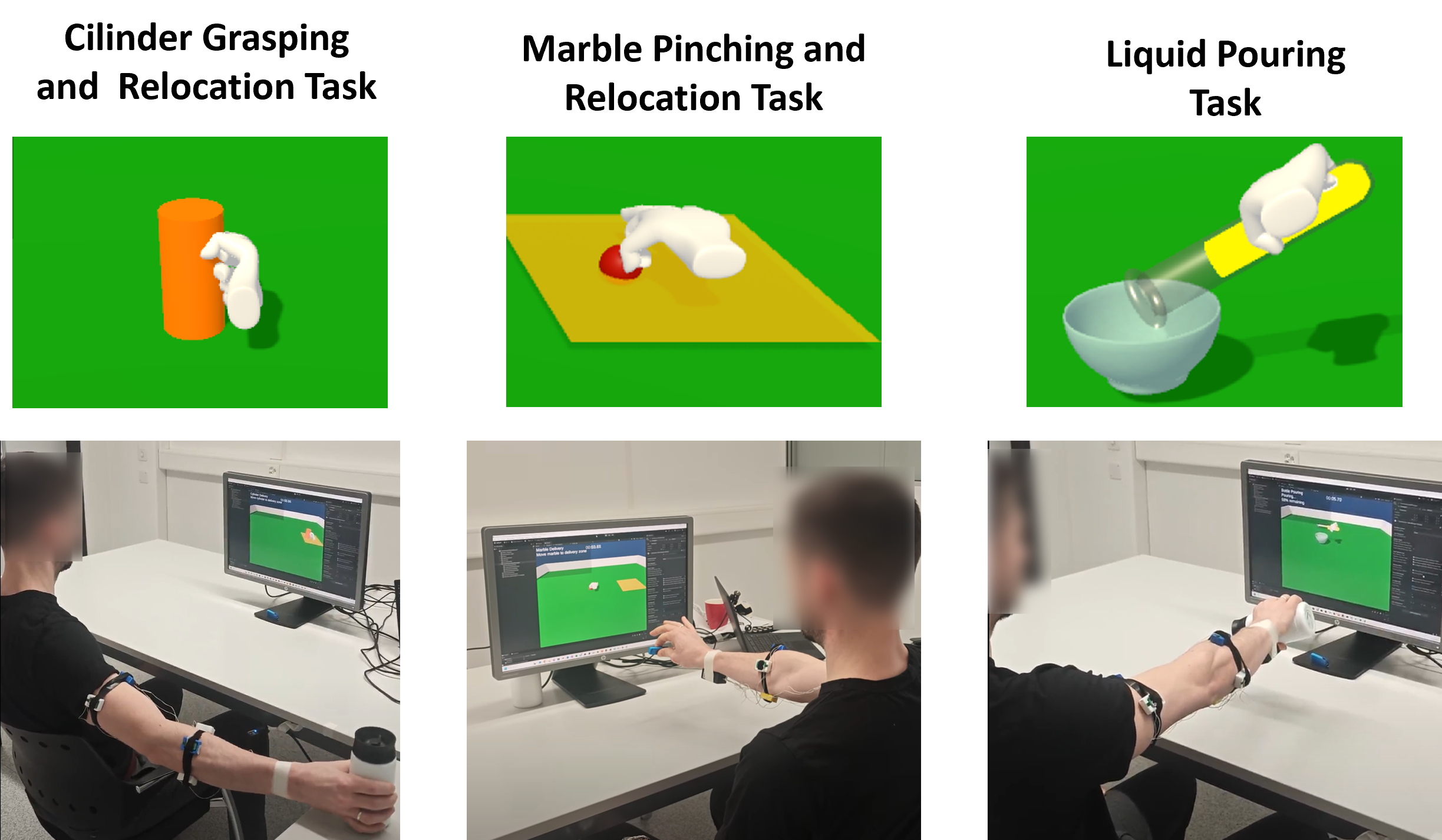}
    \vspace{-0.2cm}
    \caption{Top: visualization of the Unity-based interaction environment used for the three experimental tasks. Bottom: Example of a participant interacting with the proposed interface during task execution.}
    \vspace{-0.5cm}
    \label{fig:task}
\end{figure}

\vspace{-0.1cm}
We collect data from five subjects performing three functional tasks (Fig.~\ref{fig:task}), inspired by \cite{sgambato_2025_virtual}: cylinder relocation, marble pinching, and liquid pouring.
We collect data in sessions, over multiple days, with sensor repositioning between sessions. In each session, subjects perform five repetitions of each task. For each task, subjects are seated upright in front of a monitor, with the BioGUI displaying the task instructions and the timing cues. 
Each task starts from the rest position, with the elbow flexed at $90^\circ$, the upper arm close to the trunk, and the forearm in neutral rotation, mapping the reference position previously described (see Sec. \ref{subsec:software_architecture} - Rotation Tracking). 
Tasks progress as follows:
 
\begin{figure}[h]
    \centering
    \vspace{-0.2cm}
    \includegraphics[width=0.8\linewidth]{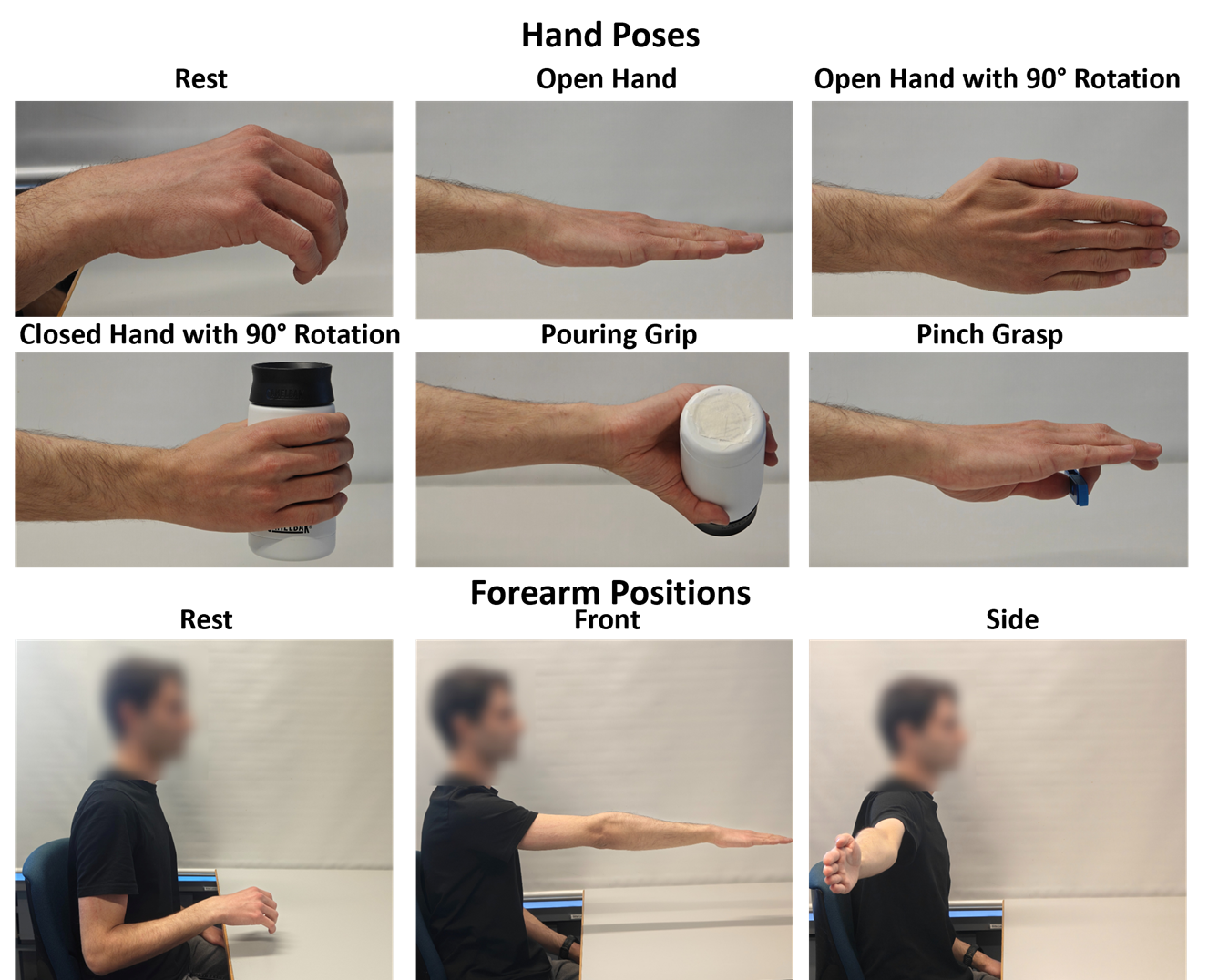}
    \vspace{-0.2cm}
    \caption{Overview of Hand Poses (top) and Forearm Positions (bottom) included in the data collection protocol}
    \label{fig:pose_positions}
    \vspace{-0.4cm}
\end{figure}


 \textbf{\textit{i)} Cylinder Grasping and Relocation:} the arm is advanced toward a cylindrical object (e.g., cup), while the hand opens and the forearm undergoes a $90^\circ$ rotation to align the palm vertically with the cylinder; then, the hand grasps and lifts the cylinder; while maintaining the grasp, the arm is abducted laterally by $90^\circ$ to transfer the object to the side; finally, the cylinder is released at the target position by reopening the hand while maintaining forearm rotation.

 \textbf{\textit{ii)} Marble Pinching and Relocation:} the arm is moved forward to reach an object (e.g., marble, USB stick), while the hand opens while maintaining the calibrated horizontal orientation (palm-down); then, a pinch (thumb--index opposition) is performed to grasp the small object; while maintaining the pinch, the arm is moved laterally by $90^\circ$ to relocate the object to the side; finally, the object is placed at the target location, and the hand opens to release it.

\textbf{\textit{iii)} Liquid pouring:} the arm is moved forward to reach the object (a bottle), while the hand opens and the forearm undergoes a $90^\circ$ rotation to align the palm vertically with the object; the hand grasps and lifts the bottle; finally, the forearm rotates to tilt the bottle, reproducing a pouring task.

Each position within a task is held for $5\,\mathrm{s}$.
The three tasks define a total of 6 distinct \textit{hand poses} and 3 \textit{forearm positions} in the 2D space, as shown in Fig. \ref{fig:pose_positions}. 
\vspace{-0.3cm}
\subsection{Network Architecture and Offline Model 
Training}\label{subsec:architecture}
\vspace{-0.1cm}
\input{tables/architecture}
We propose a CNN-based architecture using \ac{US} and \ac{ACC} data. The same architectural backbone is used to discriminate the 6 \textit{hand poses} and the 3 \textit{forearm positions} (to control positioning in the 2D space).
The model processes the two modalities through a late-fusion approach. It consists of a \textit{\ac{US} feature extractor}, followed by \textit{fully connected layers} where \ac{US} features are fused with the \ac{ACC} data.

The US input consists of $C=6$ transducers, with $T=397$ depth samples per window, arranged as $(B, C, D)$, where $B$ denotes the batch size. Prior to convolution, the input is reshaped to $(B, 1, D, C)$.
The \textit{US Feature Extractor} is composed of two convolutional blocks. Each block consists of a Conv2D layer followed by batch normalization, ReLU activation, dropout regularization (dropout rate set to 0.05), and max pooling in the depth dimension. The convolution kernels have size $(7 \times 1)$, enabling feature extraction along the depth dimension; max pooling with a kernel $(4 \times 1)$ is used. After the second block, the resulting \ac{US} feature map is flattened to form a 144-dimensional US feature vector.

A vector of three \ac{ACC} features is concatenated with the flattened \ac{US} representation, producing a joint feature vector. This fused representation is processed by a fully connected classification head composed of two hidden layers with ReLU activation, followed by the final output layer. The output layer contains $N$ classes, with $N=6$ for the hand pose model and $N=3$ for the forearm position model. The model features 13710 and 13599 parameters for the hand pose and forearm pose models, respectively. 

We use the cross-entropy loss function and the Adam optimizer with a learning rate of $10^{-3}$. A maximum of 50 epochs is considered; early stopping is applied based on the validation loss, with training stopped after 5 consecutive epochs without improvement.
For offline training, each dataset sample is formed by collecting the most recent waveform from each of the 6 \ac{US} transducers, yielding an input of size $6\times397$, and by averaging the 3 \ac{ACC} channels over the same rolling window. With \ac{US} transducers excited in a round-robin fashion, consecutive samples differed only in the latest acquired US waveform. 
To account for subject reaction time, $1\text{s}$ of data is removed at each transition between hand-forearm states.

\vspace{-0.2cm}
\subsection{Online Data Evaluation Methodology}
\vspace{-0.1cm}

We perform online validation with all five subjects on the same three functional tasks (Sec.~\ref{subsec:offline_datacollection}). 
\textit{Hand poses} and \textit{forearm positions} are controlled by their respective models (see \review{Sec.} \ref{subsec:architecture}).
The \textit{Hand Rotation} is estimated using the analytic acceleration-based model described in Sec.~\ref{subsec:software_architecture}. The interaction logic between the subject, the virtual hand, and the virtual objects was implemented in Unity. 

For the \textit{hand pose}, the Unity application accepts four hand poses: Rest, Open, Closed, and Pinch. Since the classifier distinguishes six classes (Hand Open, Hand Open with $90^\circ$ rotation, Hand Closed with $90^\circ$ rotation, Pinch, Pouring, and Rest), a mapping is applied: Hand Open and Hand Open with $90^\circ$ rotation are aggregated into Open, Hand Closed with $90^\circ$ rotation and Pouring are aggregated into Closed. The remaining classes (Rest, Pinch) are forwarded unchanged. 

For the \textit{forearm position}, the classifier distinguishes three classes (Rest, Forward, Side), which are forwarded directly to Unity without aggregation. 

All three control signals (hand pose, forearm position, and hand rotation) are updated at $30~\text{Hz}$, corresponding to each new set of A-mode \ac{US} and \ac{ACC} samples. To improve robustness, the pose and position predictions are each smoothed independently using a majority vote over 30 consecutive predictions.

For the online validation, each subject first collected an additional acquisition session, following the same protocol described in Sec.~\ref{subsec:offline_datacollection}, for subject-specific fine-tuning. Specifically, five fine-tuned models are trained per subject,  using one to five repetitions from the newly collected session, allowing the assessment of the impact of different amounts of fine-tuning data.
Fine-tuning is performed with a learning rate of $10^{-3}$, using 70\% of the data for training and 30\% for validation. We apply early stopping with a patience of 5 epochs and train for a maximum of 30 epochs. In addition, a zero-shot model without fine-tuning is evaluated. Overall, this results in 6 models per subject for the online validation.

Before the evaluation, each subject is given 5 to 10 minutes to freely interact with the Unity environment and try the tasks using the model fine-tuned with 5 repetitions. The online validation then starts. The models are used in randomized order, unknown to the subjects.  For each model, the subjects complete 5 trials per task. The tasks displayed in the Unity engine map the ones collected during the Offline Data Collection. After each completed or failed trial, a countdown of 5\,s is displayed before the start of the next trial.
If a trial is not completed within 30\,s, it is considered failed and stopped. \review{Recovery attempts are permitted during a trial (e.g., retrieving a dropped box) and remain successful if the final objective is met.} Relocation tasks are considered successful if the object is placed within the delivery zone; the pouring task is considered successful if the bottle is emptied. We use \ac{CT} and \ac{SR} as evaluation metrics, retrieved from the Unity engine and saved to a \texttt{.csv} file. \ac{SR} is defined as the percentage of successful trials over all attempts for a given condition. \ac{CT} is computed over successful trials only and reported as mean \(\pm\) standard deviation across subjects.

%% file: tables/architecture.tex
\begin{table}[t]
\centering
\caption{Architecture of the proposed US+ACC CNN model.}
\label{tab:us_imu_cnn_architecture}
\scriptsize
\setlength{\tabcolsep}{4pt}
\begin{tabular}{c c c c c}
\hline
\textbf{Type} & \textbf{Layer} & \textbf{\#Filt.} & \textbf{Kernel} & \textbf{Output} \\
\hline
Input & US input & -- & -- & $(1,397,6)$ \\
\hline
$\phi^{1}_{US}$ & Conv2D & 1 & $(7\times1)$ & $(1,397,6)$ \\
                & MaxPool2D & -- & $(4\times1)$ & $(1,99,6)$ \\
\hline
$\phi^{2}_{US}$ & Conv2D & 1 & $(7\times1)$ & $(1,99,6)$ \\
                & MaxPool2D & -- & $(4\times1)$ & $(1,24,6)$ \\
\hline
$\phi^{3}_{US}$ & Flattened US & -- & -- & $US_{feat}$ $(144$) \\
\hline\hline
Input & ACC input & -- & -- &  $ACC_{in}$ ($3$) \\
\hline
Fusion & $[US_{feat}, ACC_{in}]$ & -- & -- & $147$ \\
\hline
$\phi^{4}_{US+ACC}$ & Dense + ReLU & -- & -- & $73$ \\
\hline
$\phi^{5}_{US+ACC}$ & Dense + ReLU & -- & -- & $36$ \\
\hline
$\phi^{6}_{US+ACC}$ & Dense & -- & -- & $N$ \\
\hline
\end{tabular}
\begin{tablenotes}
\footnotesize
\item US data are reshaped from $(B,6,397)$ to $(B,1,397,6)$ before convolution. Convolution uses the "same" padding; pooling is applied only along the depth dimension. ACC data are fused after the CNN encoder through concatenation. The same architecture is used for both the hand pose model and the forearm position model, differing only in the output layer size, with $N=6$ classes for hand pose classification and $N=3$ classes for forearm position classification.
\end{tablenotes}
\vspace{-0.5cm}
\end{table}

%% file: sections/4_Results.tex
\vspace{-0.2cm}
\section{Results}
\label{sec:Results}
\vspace{-0.2cm}
\subsection{Offline Results}
\begin{figure}[!htbp]
    \centering
    \includegraphics[width=0.7\linewidth]{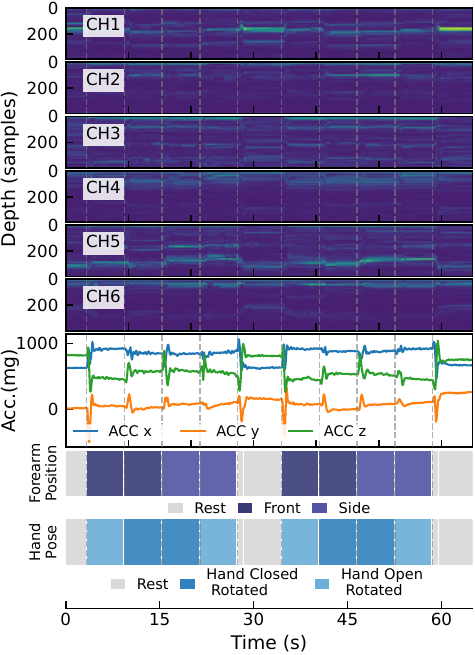}
    \vspace{-0.3cm}
    \caption{Example sequence of the data collection with 6 channels of US data, acceleration, and labels for forearm position and hand pose models for the Cylinder Grasping and Relocation Task.}
    \label{fig:res_m_mode}
    \vspace{-0.8cm}
\end{figure}

\input{tables/results_acc}
Figure \ref{fig:res_m_mode} shows a representative image of \ac{US} data, \ac{ACC}, and corresponding forearm position hand pose labels for one of the three tasks (Cylinder Grasping and Relocation), and Table \ref{tab:avg_acc} reports the average results for each subject, for the hand pose and forearm position models, respectively, \review{comparing the performance with and without the use of ACC data. For both hand and position models, including ACC data improves the performance.}
For the hand pose models, we achieve an average inter-session accuracy among the 5 subjects of $80\pm6\%$. 
The main misclassification occurs between the \textit{Open} and \textit{Pinch} gestures (not shown).  We attribute this behavior to two main factors: the biomechanical similarity of the two movements and the limited number of transducers in the sensing setup. Since these gestures differ mainly in thumb opposition and index finger flexion, they likely generate similar muscle deformation patterns and, consequently, similar recorded signals. Moreover, only two transducers were positioned over the forearm flexor region, reducing the spatial selectivity available to distinguish them, with inter-session armband repositioning further amplifying this issue by altering the measured deformation patterns. A second recurrent error is between \textit{Hand Open-Rotated} and \textit{Hand Closed-Rotated}, which we attribute to the similar forearm orientation of these gestures: the signal features associated with the supinated position may partially mask the differences arising from finger flexion. In this case, it should also be noted that subjects were not asked to fully close the hand, but rather to grasp an object. As a result, the fingers were not completely flexed, making the distinction between the two rotated gestures even less pronounced.

\begin{figure}[h]
    \centering
    \includegraphics[width=0.8\linewidth]{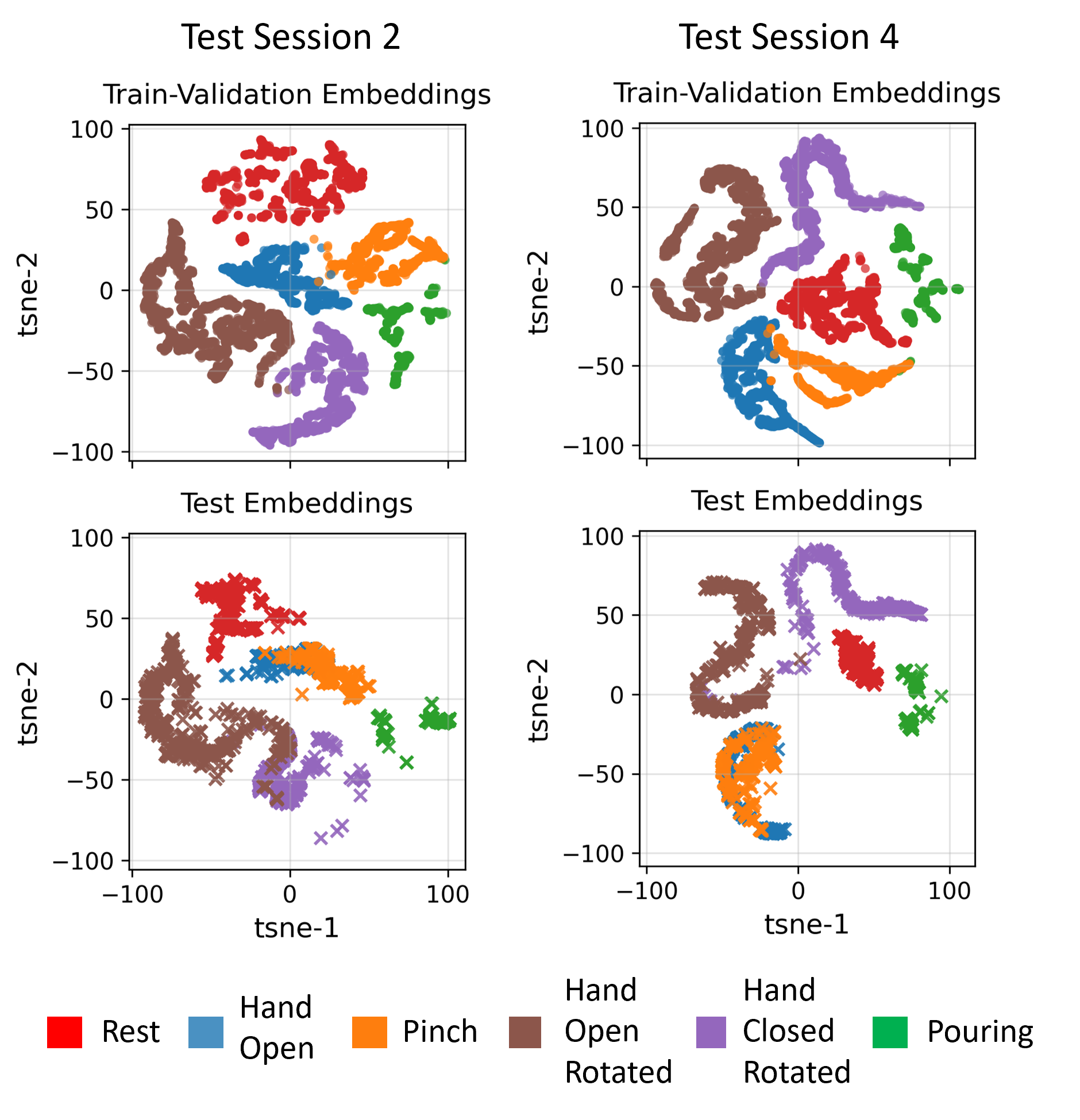}
    \vspace{-0.3cm}
    \caption{\ac{TSNE} projections of the latent representations before the classification output for the hand pose model for one subject (S05), for two representative inter-session folds. Left: model trained on sessions 1,3,4,5, tested on session 2; Right: model trained on sessions 1,2,3,5, tested on session 4. For both models, the top row shows the training-validation embeddings and the bottom row the test embeddings.}
    \vspace{-0.6cm}
    \label{fig:tsne_cms}
\end{figure}
Figure~\ref{fig:tsne_cms} shows that inter-session errors are largely session-dependent, likely due to armband repositioning and slight differences in movement execution across sessions. We report the \ac{TSNE} projections of the latent features before classification for the hand pose model of subject S05 in two inter-session folds. In the first fold, the model is trained on sessions 1, 3, 4, and 5 and tested on session 2; in the second fold, it is trained on sessions 1, 2, 3, and 5 and tested on session 4. The top row shows train-validation embeddings, and the bottom row shows test embeddings. In the first fold (left column, 75\% accuracy), the lower performance is associated with a partial migration of the test embeddings across classes: \textit{Hand Open, Rotated} shifts toward \textit{Hand Closed, Rotated}, and vice versa, while \textit{Pinch} shifts toward the \textit{Hand Open} cluster. In Fold~2 (88\% accuracy), \textit{Hand Open, Rotated} and \textit{Hand Closed, Rotated} remain more stable, with limited class migration. Although a visible shift between \textit{Hand Open} and \textit{Pinch} is still present, the class structure is better preserved, resulting in fewer errors.

Overall, these findings highlight the importance of incorporating session-specific recalibration data to mitigate the impact of inter-session armband repositioning on model performance.

\vspace{-0.2cm}
\subsection{Online Results}
\vspace{-0.1cm}

\begin{figure}[t]
    \centering
\includegraphics[width=0.85\linewidth]{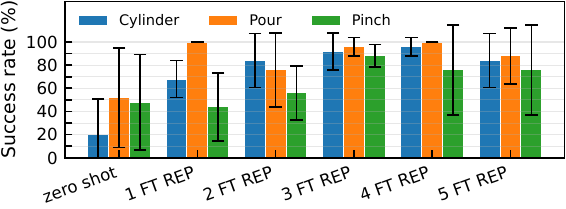}
    \vspace{-0.2cm}
    \caption{Average \ac{SR} per task in the online validation with different amounts of fine-tuning data.}
    \vspace{-0.2cm}
    \label{fig:res_success_rate}
\end{figure}

Figure~\ref{fig:res_success_rate} reports the mean and standard deviation of the task \ac{SR} across subjects for each evaluated model (zero-shot and fine-tuning using 1--5 repetitions). The best overall performance was obtained with fine-tuning on 3 repetitions (3 FT REP). For this setting, the \ac{SR} reached \(92.0 \pm 16.0\%\) for Cylinder Grasping and Relocation, \(96.0 \pm 8.0\%\) for Liquid Pouring, and \(88.0 \pm 9.8\%\) for Marble Pinching and Relocation. These results confirm that limited subject-specific fine-tuning (5 min) is sufficient to substantially improve online control.

Across all tasks, \acp{SR} increased with additional fine-tuning. This trend is most pronounced for the Cylinder Grasping and Relocation task, which improves from $\approx$20\% in the zero-shot setting to nearly 100\% when fine-tuning on 3-4 repetitions. The Liquid Pouring task benefits already from limited fine-tuning, suggesting that the associated gross grasping motion and the characteristic \ac{ACC} patterns during rotation are comparatively easy to separate from other classes. This is consistent with the trends observed in the offline inter-session results (see Fig. \ref{fig:tsne_cms}), where pouring is always clearly clustered compared to other gestures. In contrast, the Marble Pinching and Relocation task shows a smaller gain, improving from around 50\% (zero-shot and 1--2 repetitions) to 70--80\% (3--5 repetitions), likely due to the need to detect a fine pinch gesture while the arm simultaneously transitions from the front to the side position, increasing ambiguity between different states.

\begin{figure}[t]
    \centering
    \includegraphics[width=0.85\linewidth]{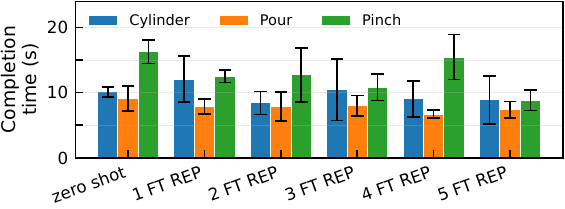}
    \vspace{-0.2cm}
    \caption{Average \ac{CT} per task in the online validation with different amounts of fine-tuning data.}
    \vspace{-0.6cm}
    \label{fig:res_completion_time}
\end{figure}

Figure~\ref{fig:res_completion_time} shows the mean and standard deviation of \ac{CT} across subjects, computed over successful trials only. The best \review{result is obtained with} 3 FT REP, \review{with} a \ac{CT} of \(10.47 \pm 4.73\) s for Cylinder Grasping and Relocation, \(8.00 \pm 1.56\) s for Liquid Pouring, and \(10.83 \pm 2.02\) s for Marble Pinching and Relocation. 
For the Cylinder Grasping and Relocation task and the Liquid Pouring task, \acp{CT} are relatively consistent across models (approximately 7--12\,s). Pinch relocation exhibits larger variability and generally longer \acp{CT} (approximately 10--20\,s), consistent with the increased difficulty of reliably detecting and maintaining the pinch state during arm motion.

During the online validation, we observed the following qualitative behaviors. Pose predictions (e.g., hand open, closed, pinch) were stable when the forearm position was held static (e.g., front). Fine pose states (marble pinching) were less reliable than gross pose states (cylinder grasping). Pose stability decreased during dynamic transitions of the forearm position (front to side), leading to a common failure mode in which objects (marble or cylinder) were unintentionally dropped while being transported. For some configurations (mostly zero-shot and occasionally fine-tuned models), subjects experienced repeated failures due to consistent error modes (e.g., inability to trigger pinch pickup or dropping the object during pouring), which contributed to the larger standard deviations in the \ac{SR} in those settings.

In terms of power consumption, our solution consumes $19.9~\textrm{mW}$, as measured with a Power Profiler from Nordic Semiconductor~\cite{nordicPower}, thereby enabling continuous use for more than 2.5 days on a small 350~mAh LiPo battery.
\vspace{-0.2cm}
\subsection{Comparison with Previous Works}
\label{soa:comparison}
\input{tables/related_new}
Table \ref{tab:us_comparison_transposed} reports a comparison with previous works.
From a system-level perspective, our solution is, to the best of our knowledge, the only one enabling truly wearable interaction with \ac{VR} environments. Compared with prior systems, it features a smaller and lighter form factor, and is the only one with wireless connectivity. Also, our approach avoids the need for standard ultrasonic gel, enabling fully dry acquisition, which was previously explored only in \cite{sgambato_2025_virtual}. Moreover, all studies that explicitly report power consumption rely on systems with power consumption exceeding $>1 \textrm{W}$. In contrast, the proposed WULPUS-based solution operates below 20 mW, corresponding to more than one order of magnitude reduction in power consumption. This efficiency enables over 2.5 days of continuous operation with a \(350\,\text{mAh}\) LiPo battery. Finally, this work is also, to the best of our knowledge, the first wearable system for forearm and upper arm monitoring combining \ac{US} and \ac{ACC} data.

From an application perspective, we provide a qualitative evaluation with respect to prior works, as a quantitative comparison is not feasible due to substantial differences in task complexity (Table \ref{tab:us_comparison_transposed}). Early works (\cite{lu_2022}) focused on simple gesture classification without any functional interaction, remaining far from realistic conditions. The \ac{TAC} tests \cite{yang_2022_sonomyographic,sgambato_2023} represent a step toward functional evaluation, but constrain the hand to a 2D space and do not include interaction with virtual objects. The proposed framework is aligned with \cite{sgambato_2025_virtual}, extending it toward more wearable and functionally complete \ac{VR} interaction. As in \cite{sgambato_2025_virtual}, we enable object manipulation tasks such as liquid pouring and cylinder relocation, while additionally introducing pinch grasping, covering both gross and fine motor control. In contrast to \cite{sgambato_2025_virtual}, where hand position and orientation are obtained via an external optical tracking system, our approach relies entirely on wearable sensing.
Although 2D hand and arm position control using \ac{US} has also been explored in \cite{tang_2025_synchronous}, that system relies on non-wearable acquisition electronics, limiting practical applicability. A direct quantitative comparison with \cite{sgambato_2025_virtual} is not possible, as performance is reported as correlation with motion-capture angles rather than task-level metrics such as \ac{CT} or \ac{SR}. 

%% file: tables/results_acc.tex


\begin{table}[t]
\centering
\caption{Average balanced accuracy across subjects for multimodal (US+ACC) and \review{US-only} networks.}
\label{tab:avg_acc}
\scriptsize
\setlength{\tabcolsep}{3pt}
\renewcommand{\arraystretch}{1.1}
\begin{tabular}{lcc!{\vrule width 1.2pt}cc}
\toprule
\textbf{Subject}
& \multicolumn{2}{c}{\textbf{Hand pose}}
& \multicolumn{2}{c}{\textbf{Forearm position}} \\
\cmidrule(r){2-3}\cmidrule(l){4-5}
& \textbf{US+ACC}
& \review{\textbf{US-only}}
& \textbf{US+ACC}
& \review{\textbf{US-only}} \\
\midrule
S1  & 0.76$\pm$0.07 & \review{0.57$\pm$0.19} & 0.74$\pm$0.15 & \review{0.70$\pm$0.18} \\

S2  & 0.70$\pm$0.09 & \review{0.76$\pm$0.13} & 0.72$\pm$0.14 & \review{0.70$\pm$0.10} \\

S3  & 0.87$\pm$0.05 & \review{0.72$\pm$0.12} & 0.91$\pm$0.04 & \review{0.77$\pm$0.03} \\

S4  & 0.85$\pm$0.04 & \review{0.70$\pm$0.04} & 0.72$\pm$0.11 & \review{0.82$\pm$0.04} \\

S5  & 0.86$\pm$0.07 & \review{0.67$\pm$0.10} & 0.77$\pm$0.16 & \review{0.80$\pm$0.11} \\

\midrule
\textbf{Avg}
& \textbf{0.80$\pm$0.06} & \review{\textbf{0.69$\pm$0.07}}
& \textbf{0.77$\pm$0.07} & \review{\textbf{0.76$\pm$0.06}} \\
\bottomrule
\end{tabular}
\vspace{-0.5cm}
\end{table}

%% file: tables/related_new.tex
\setlength{\lightrulewidth}{0.2pt}
\begin{table*}[t]
\centering
\scriptsize
\caption{Comparison of ultrasound-based hand and forearm interaction systems at the system and application level.}
\vspace{-0.3cm}
\label{tab:us_comparison_transposed}
\setlength{\tabcolsep}{1.8pt}
\renewcommand{\arraystretch}{0.5}

\begin{tabularx}{\textwidth}{@{}>{\centering\arraybackslash}p{0.33cm} p{1.55cm} *{6}{Y}@{}}
\toprule
\textbf{} & \textbf{}
& Lu, 2022 \cite{lu_2022}
& Yang, 2023 \cite{yang_2022_sonomyographic}
& Sgambato, 2023 \cite{sgambato_2023}
& Tang, 2025 \cite{tang_2025_synchronous}
& Sgambato, 2025 \cite{sgambato_2025_virtual}
& \textbf{Our Work} \\
\toprule

\multirow{8}{=}{\raisebox{-8ex}[0pt][0pt]{\rotatebox[origin=c]{90}{\textbf{System}}}}
& Platform
& ELONXI
& Custom \cite{yang_2021_wearable}
& MOUSE \cite{mouse_fournelle_2019}
& Benchtop devices
& MOUSE \cite{mouse_fournelle_2019}
& WULPUS \cite{frey_wulpus_2022} \\
\cmidrule(lr){2-8}

& Size [mm]
& NA
& $132\times90\times30$
& $184\times123\times33$
& NA
& $184\times123\times33$
& $31\times54\times26$ \\
\cmidrule(lr){2-8}

& Weight [g]
& NA
& 190
& 610 
& NA
& 610 
& 34 \\
\cmidrule(lr){2-8}




& PRF [Hz]
& 10
& 10
& 108
& 72
& $377.6\pm16$  
& $30^{a}$ \\
\cmidrule(lr){2-8}

& Connectivity
& Cabled
& Ethernet/WiFi
& USB 3.0
& Cabled
& USB 3.0
& BLE \\
\cmidrule(lr){2-8}

& Wearability
& No
& No 
& No
& No
& No
& Yes\\
\cmidrule(lr){2-8} 

& Power [mW]
& NA
& 3500
& 12000
& NA
& 12000
& 19.9\\

\toprule

\multirow{10}{=}{\raisebox{-10ex}[0pt][0pt]{\rotatebox[origin=c]{90}{\textbf{Application}}}}
& \# Transducers
& 4
& 8
& 24
& 36
& 32
& 4 + 2 \\
\cmidrule(lr){2-8}

& Placement
& Forearm
& Forearm 
& Forearm 
& Forearm, upper arm, chest 
& Forearm 
& Forearm, upper arm \\ 
\cmidrule(lr){2-8}


& Coupling
& Gel
& Gel
& Gel
& Silicone + Gel
& Dry-contact silicone + moisturizer
& Hydrogel pads \\
\cmidrule(lr){2-8}

& Modalities
& US
& US
& US
& US
& US
& US + \ac{ACC} \\
\cmidrule(lr){2-8}

& Live training
& From scratch
& From scratch
& From scratch
& NA
& No fine-tuning
& Fine-tuning\\
\cmidrule(lr){2-8}


& Live task
& Gesture recognition
& TAC
& TAC
& 1 functional task
& 5 functional tasks
& 3 functional tasks \\
\cmidrule(lr){2-8}

& Controlled part
& Hand
& Hand
& Hand
& Hand + Arm
& Hand
& Hand + forearm position (2D space) \\
\cmidrule(lr){2-8}

& Problem type
& Class, 10 gestures
& Reg, 3 DoF
& Reg, 3 DoF 
& Class (Hand \ac{OC}) + Reg, 3 DoF
& Reg, 4 DoF
& Class, 6 Hand Poses + 3 forearm positions\\
\cmidrule(lr){2-8}

& Online metrics
& MST, CT, CR, RA
& CR, CT, SE
& CR, CT
& NA
& Correlation with optical tracking
& CR, CT \\


\bottomrule
\end{tabularx}

\vspace{0.3em}
\begin{flushleft}
\vspace{-0.2cm}
\footnotesize
$^{a}$ Time-multiplexed acquisition; effective per-transducer rate is $30/N_{\mathrm{transducers}}$. 
Class: classification; Reg: regression, MST: Motion Selection Time; CR: Completion Rate; RA: real-time accuracy, also defined as SR: success rate; SE: stability error
\vspace{-0.6cm}
\end{flushleft}
\end{table*}

%% file: sections/5_Conclusions.tex
\vspace{-0.2cm}
\section{Conclusion}\label{sec:conclusions}
\vspace{-0.1cm}

In this work, we presented a fully wearable multimodal sensing system combining \ac{US} and \ac{ACC} for concurrent monitoring of the forearm and upper arm using the WULPUS platform. We present an open-source, end-to-end real-time framework for data acquisition and interaction with \ac{VR} environments, enabling reproducible evaluation across sensing platforms and algorithms. Within this framework, we demonstrated three functional tasks involving both gross and fine object manipulation and relocation. A multimodal learning pipeline was introduced for simultaneous hand pose and 2D forearm position estimation, achieving average offline inter-session accuracies of \(80\pm6\%\) and \(77\pm7\%\) across five subjects, respectively. Real-time validation showed that, even after sensor repositioning, performance can be recovered with minimal fine-tuning: with only 3 repetitions, the system achieved task \ac{SR} of \(92.0\pm16.0\%\), \(96.0\pm8.0\%\), and \(88.0\pm9.8\%\), with corresponding \ac{CT} of \(10.47\pm4.73\)\,s, \(8.00\pm1.56\)\,s, and \(10.83\pm2.02\)\,s, for cylinder grasping/relocation, liquid pouring, and marble pinching/relocation, respectively.

A limitation of this work is that hand pose and forearm position are modeled as discrete states. In addition, online predictions are smoothed through a majority vote over 30 frames, yielding an effective latency of $\approx$ \(1\,\mathrm{s}\). While this reduces responsiveness, it suppresses transient misclassifications and improves stability. 
Future work will investigate proportional and regression-based interaction for smoother and lower-latency control. 
\review{In addition, future work will extend the evaluation to a larger cohort of subjects, including variability in the demographics, as well as extending the number of tasks. }
Another limitation is the limited number of sensing units. Future developments will target higher channel count and increased wireless bandwidth (e.g., via Wi-Fi) for wider body coverage. Additional sensing locations, such as the wrist and shoulder, as well as multimodal fusion with other signals such as \ac{EMG}, will also be explored.